\documentclass[prb,preprint,aps,showpacs,longbibliography,
lengthcheck,superscriptaddress]{revtex4-1}

\usepackage{graphicx}%
\usepackage{dcolumn}
\usepackage{amssymb}
\usepackage[usenames,dvipsnames,svgnames,table]{xcolor}

\newcommand{\sinp}{Saha Institute of Nuclear Physics, I/AF Bidhannagar, Kolkata 700 064, India}

\begin{document}

\title{Universal scaling in disordered systems and non-universal exponents}

\author{K. K. Bardhan}
\affiliation{Kalpana Chawla Centre  for Space and Nano Sciences, 3F Swamiji Nagar, Kolkata 700 030, India}
\email{kamalk.bardhan@saha.ac.in}

\author{D. Talukdar}
\affiliation{\sinp}
\thanks{Present Address: Division of Physics and Applied Physics, School of Physical and Mathematical Sciences, Nanyang Technological University, Singapore 637371}
\email{dtalukdar@ntu.edu.sg}

\author{U. N. Nandi}
\affiliation{Department of Physics, Scottish Church College, 1 $\&$ 3 Urquhart Square, Kolkata-700 006, India}

\author{C.D. Mukherjee}
\affiliation{\sinp}


\begin{abstract} 
The effect of an electric field on conduction in a disordered system is an old but largely unsolved problem. Experiments cover an wide variety of systems - amorphous/doped semiconductors, conducting polymers, organic crystals, manganites, composites, metallic alloys, double perovskites - ranging from strongly localized systems to weakly localized ones, from strongly correlated ones to weakly correlated ones. Theories have singularly failed to predict any universal trend resulting in separate theories for separate systems. Here we discuss an one-parameter scaling that has recently been found to give a systematic account of the field-dependent conductance in two diverse, strongly localized systems of conducting polymers and manganites. The nonlinearity exponent, \textit{x} associated with the scaling was found to be nonuniversal and exhibits structure. For two-dimensional (2D) weakly localized systems, the nonlinearity exponent \textbf{\textit{x}} is $\geqslant 7$ and is roughly inversely proportional to the sheet resistance. Existing theories of weak localization prove to be adequate and a complete scaling function is derived. In a 2D strongly localized system a temperature-induced scaling-nonscaling transition (SNST) is revealed. For three-dimensional (3D) strongly localized systems the exponent lies between -1 and 1, and surprisingly is quantized (\textit{x} $\approx$ 0.08 \textit{n}). This poses a serious theoretical challenge. Various results are compared with predictions of the existing theories.
\end{abstract}

 
\maketitle

\section{Introduction}
Effects of disorder on transport phenomena have been studied for more than four decades\cite {mottbook}, yet understanding is still far from complete. One of the effects of disorder is to drive a sample into a nonOhmic regime even under small bias. This effect is particularly pronounced in small-sized samples such as nano- or micro-devices. Apart from physical reasons it is thus necessary to understand the phenomena of nonlinear transport to ensure that the devices operate as desired with optimum properties. In the early period the focus was on the materials such as amorphous/doped semiconductors, glasses which are recognized today as strongly localized (SL) systems characterized by the exponential dependence of conductivity on temperature. In many such systems the conduction occurs through variable range hopping (VRH) between localized states randomly occurring in space with the conductivity given by\cite {mottbook}
\begin{equation}
\sigma_o(T) = \sigma_{oo} \exp \left [ -{\left ( {T_o} \over T \right )}^m \right ].
\label{eq:mott}
\end{equation}
Here $\sigma_{oo}$ is a prefactor, $T_o$ a characteristic temperature, and the exponent \textit{m} lies in the range $1/4 \leq m < 1$. \textit{m}=1 corresponds to nearest neighbor hopping.  Inclusion of electron-electron interactions leads to \textit{m}=1/2\cite {efros75}. $T_o$ is a decreasing function of the localization length $\xi$. Later new disordered regimes, namely weakly localized (WL) metallic state in 2D and metal-insulator transition (MIT) in 3D were identified\cite {abrahams79}. In contrast to Eq. (\ref{eq:mott}), a WL sample was predicted to have a logarithmic variation of the Ohmic conductivity with temperature. The disorder-induced WL-SL crossover in 2D also became a topic of intense activities\cite {abrahams01}. The MIT was probed by varying disorder with doping and the results were analyzed using phenomenological scaling concepts\cite {sondhi97} borrowed from theories of general critical phenomena.

Introduction of an electric field helps gather information which are not available from temperature study alone as it involves features of the conduction mechanism not considered in the Ohmic regime (some examples can be found below). The effects of application of an field upon a WL system result in an effective electron (or, hole) temperature $T_e$ above the phonon bath temperature \textit{T} of the substrate. The hot electron model in the WL regime\cite {anderson79} (HEM-WL) does not involve any exchange of energy between the sample and the substrate at low temperatures but explains the characteristic logarithmic field-dependence of conductance. The electron heating effects have been shown to account for the nonlinear conductance also in the SL regime of a 2D system\cite {gershen00,leturcq03} but with a provision of exchanging energy between the sample and the substrate. The nonOhmic resistance in the SL regime is still described by the equation (Eq. \ref{eq:mott}) for Ohmic resistance but with \textit{T} replaced by an effective temperature $T_e$ given by the following energy balance relation
\begin{equation}
V^2 /R = C ({T_e}^{\beta} - T^{\beta}),
\label{eq:hemSL}
\end{equation}
where \textit{R} is the sample resistance, \textit{V} the applied bias, \textit{C} a constant incorporating electron-phonon interactions and $\beta$ an exponent. Henceforth, the model will be referred as HEM-SL.

The situation in 3D SL regime is still theoretically inconclusive as none of the proposed models gives a consistent and adequate explanation of the relatively abundant data available in the literatures. The present models are basically of two types\cite {[{For a recent review see }]caravaca10} - so called 'field-effect' models and hot electron model (HEM-SL). According to the field-effect models\cite {shklov76, pollak76} which are essentially based upon percolation among localized carriers the conductivity for intermediate fields is given by
\begin{equation}
\sigma(T,F) = \sigma_o(T) \; \exp (F/F_l).
\label{eq:lowF}
\end{equation}
Here $\sigma_o(T)=\sigma(T,0)$ is the zero-field or Ohmic conductivity which in case of VRH is given by Eq. (\ref{eq:mott}). The field scale $F_l$ is given by
\begin{equation}
F_l = k_B T/eL_h,
\label{eq:lfscale}
\end{equation}
where $L_h$ is a length related to the hopping length $R_h = {\xi \over 2} {(T_o/T)}^m $. In the limit of large fields, $F \ge F_u =k_B T/e \xi$) the conductivity becomes independent of temperature and is given by
\begin{equation}
\sigma \sim  {\rm{exp}} \left [- \left ( {F_u \over F} \right )^m \right ],
\label{eq:highF}
\end{equation}
with the same \textit{m} and $T_o$ as in Eq. (\ref{eq:mott}). Eqs. (\ref{eq:lowF}) and (\ref{eq:highF}) however proved inadequate to account for the experimental data in the full range of applied field (see Ref. \onlinecite {talukdar11} for a recent critique and also Ref. \onlinecite {ladieu00}). Issues include i) inconsistent results following from application of Eq. (\ref{eq:lowF}); ii) satisfaction of  Eq. (\ref {eq:highF}) only in the cases \textit{m}=1/2; and iii) failure to account for the field scale decreasing with temperature (i.e. negative nonlinearity exponent $x_T$. see below); and iv) prediction of two field scales.

The HEM-SL in 3D is similar to the one (Eq. \ref{eq:hemSL}) used in 2D SL systems but the electron-phonon coupling in this case is basically empirical\cite {wang90,zhang98,galea07}. Compared to field-effect models the HEM-SL is a general one, independent of dimensionality and not necessarily restricted to VRH systems, and applies to the entire range of field covering both small and large limits. However there is no clarity at present about the exact conditions of its applicability in 3D. A study in doped Si and Ge\cite {zhang98,galea07} reported that for $T_o/T < 135$ the HEM provided better fits to experimental data than the field-effect ones. However there are many examples (see Ref. \onlinecite {caravaca10} and Table I) where experimental values of $T_o/T$ cover wide ranges including 135 without the HEM-SL providing any valid description. Thus it is unclear whether the ratio is really a meaningful quantity for determining the applicability of the HEM.

Here we discuss a recently proposed\cite {talukdar11}, universal response to an applied electric field in the full range by adopting a model-independent scaling approach. The strong motivation comes from the remarkable similarity in the response of different disordered regimes in diverse materials such as discussed below. The scaling\cite {talukdar11,talukdar12,varade13} that the field-dependent conductance $\Sigma(M,F)$ exhibits is given by
\begin{equation}
{\Sigma(M,F) \over \Sigma_o(M)} = \Phi \left ({F \over F_o} \right ).
\label{eq:scaling}
\end{equation}
Here \textit{F} is the applied electric field. \textit{M} stands for the control parameter which is varied to change the linear conductance $\Sigma_o(M) = \Sigma(M,0)$. Temperature is the most commonly used control parameter and disorder is an example of infrequently used ones. $\Phi$ is a scaling function. The scaling appears to hold good in all regimes of disorder except where the HEM-SL provide a valid description of $\Sigma(M,F)$ in the SL regimes. A valid description has been usually taken to be a good agreement between the data and fits to the model\cite {zhang98,galea07,gershen00}. In this scaling approach the quantity which is focused upon, and which gives a more objective and quantitative characterization of the underlying conduction process, is the field scale (or, the onset field) $F_o$ at which a sample deviates from the Ohmic behavior. In fact, $\Phi$ is  $\sim 1$ for $F \le F_o$ and increases from 1 as $F$ is increased beyond $F_o$. The latter is given by
\begin{equation}
F_o(M) \sim {\Sigma_o(M)}^{x_M},
\label{eq:fscale}
\end{equation}
where the associated exponent $x_M$ is also called the nonlinearity exponent. Indeed, this quantity proved crucial in revealing a hitherto unsuspected, temperature-induced transition within the SL regime in 2D (Section III.B). It is shown that while the scaling formally is same across the regimes the exponent $x_M$ exhibits structures and is nonuniversal. In WL regimes, $x_T$ is roughly inversely proportional to the sheet resistance.  In SL regimes in 3D and possibly, also in 2D it is surprisingly found to be quantized, $x_M \approx 0.08 n$ where \textit{n} is an integer. Moreover, contrary to what is found in thermodynamic critical phenomena, the exponent exhibits multiple values in a given system. The current theories of weak localization and regimes around the MIT support the scaling. An exact scaling function for the WL regime is derived. However the scaling remains a major theoretical challenge in the SL regimes. We limit the discussion in this paper to two and three dimensions only.

In the next section we provide heuristic general arguments for the one parameter scaling and discuss why the HEM-SL fails to exhibit the same. It is then applied to different systems in different regimes of disorder in two dimension (Section III) and three dimension (Section IV). Exponents thus found along with the scaling are then discussed in Section V with a summary in Section VI.

\section{Scaling of field-dependent conductance}
We show that the scaling equations (Eqs. \ref{eq:scaling} and \ref {eq:fscale}) are the only ones possible under the following assumptions: 

1) The field-dependent conductance possesses only one field scale. This is supported by overwhelming experimental data except where the HEM-SL holds; \\ \indent 2) The scaling function $\Phi$ is a function of the field \textit{F} and disorder i.e. $\Sigma_o(M)$ only, and not explicitly of temperature, size etc.. This is in the spirit of the original scaling theory\cite {abrahams79}. 

The field-dependent conductance can be written in general as ${\Sigma(M,F) = \Sigma_o(M)}~ \Phi (F,\Sigma_o,M)$. That the linear conductance $\Sigma_o$ must be an argument of $\Phi$ can be appreciated from the fact that the nonlinearity in response to an applied electric field arises primarily due to the presence of (quenched) disorder. A highly conducting sample exhibits hardly any nonlinearity (except for joule heating which is not of concern here). In 3D, as $\Sigma_o$ decreases (or, as disorder increases) one passes from a metallic regime to a strongly localized regime through a metal-insulator transition\cite {abrahams79}. Thus for $\Phi$ to behave appropriately as a function of disorder, it must be a function of $\Sigma_o$\cite {paper26_0} which serves to characterize disorder as the most intuitive and easily accessible experimental quantity. According to the second assumption, $\Phi$ should not have an \textit{explicit} dependence on \textit{M}. This yields ${\Sigma(M,F)/ \Sigma_o(M)} = \Phi (F,\Sigma_o(M))$ or $\Phi (F/ {\Sigma_o}^{x_M})$ for an one-parameter scaling. Here $x_M$ is a (nonlinearity) exponent (with a subscript for the control parameter) in analogy to power-laws in critical phenomena with the 'critical' conductance being equal to zero.

Compare Eqs. \ref{eq:scaling}-\ref{eq:fscale} with Eqs. \ref{eq:lowF}-\ref{eq:lfscale} respectively. Eq. (\ref {eq:scaling}) unlike Eq. (\ref{eq:lowF}) is defined for the full range of applied fields. Eq. (\ref{eq:fscale}) marks a significant departure from the current literature (e.g., Eq. \ref{eq:lfscale}) in that the field scale depends solely on the linear or Ohmic conductance, not explicitly upon temperature or any other parameter. As a result, Eq. (\ref {eq:fscale}) allows seamless consideration of disorder as the control parameter (see Fig. 6b and 7 of Ref. \onlinecite {talukdar11} for an illustration). It is seen below that such a definition facilitates an uniform description of the nonlinear transport behavior across all disorder regimes. It may be mentioned that the scaling  similar to Eqs. \ref {eq:scaling}-\ref {eq:fscale} with \textit{F} and $F_o$ replaced by the frequency and its characteristic value also describes the frequency-dependent ac-conductivity at different amplitudes\cite {bardhan92}. 

Lower bound of $x_M$: It may be obtained by considering the current scale $I_o \sim \Sigma_o F_o \sim {\Sigma_o}^{x_M +1}$. Since $I_o$ should remain finite when $\Sigma_o$ is very small it follows that $x_M \geqslant -1$\cite {paper26_4}.

\subsection{Self-consistency relation (SCR)}

At large field ($q=F/F_o \gg 1$),  the conductance often becomes 'activationless' i.e. independent of \textit{T} (therefore, $\Sigma_o(T)$) (see Fig. 1) when the control parameter is temperature\cite {paper26_1}. It is seen from Eqs. (\ref{eq:scaling}) and (\ref{eq:fscale}) that this requirement is readily ensured if $\Phi (q) \sim q^{1/x_T}$ at large \textit{q} for $x_T > 0$. Thus, the scaling predicts a power-law variation at large fields for the conductance, $\Sigma(T,F) \sim F^{z_F}$ such that for $x_T > 0$,
\begin{equation}
z_F = 1/x_T.
\label{eq:zT}
\end{equation}
The above constitutes an important self-consistency relation (SCR) as the two exponents are independently determined. For, while $z_F$ is essentially a property of a single \textit{I-V} curve, determination of $x_T$ requires multiple \textit{I-V}'s measured at various \textit{T}. The above relation can also be understood by simply writing the conductance as an interpolation of asymptotic forms at small and large \textit{F}: $\Sigma \approx \Sigma_o + b F^{1/x_T}$ where \textit{b} is a constant. This leads to $F_o \sim {\Sigma_o}^{x_T}$. The asymptotic power-law that follows from the scaling contrasts with the exponential in Eq. (\ref{eq:highF}) which is not consistent with the scaling. It must be emphasized that Eqs. (\ref{eq:fscale}) and (\ref{eq:zT}) are conditional upon the existence of the scaling in Eq. (\ref{eq:scaling}).

\subsection{Failure of scaling in the HEM-SL}
The field-dependent conductance in the HEM-SL is given by Eq. (\ref{eq:mott}) with \textit{T} replaced by $T_e$ obtained from Eq. (\ref{eq:hemSL}). Support for this derives from the fact that 'hot' electrons appear to be described by a new fermi-dirac distribution that corresponds to the temperature $T_e$. Effects due to the field obviously play out differently in this model than in tunneling or 'barrier bending' models because of the intrinsic nature of temperature appearing inside the arguments of the exponential. As a result, variation of the ohmic conductance with temperature can not be factored out of the exponential leading to failure of the one-parameter scaling as in Eq. (\ref {eq:scaling}) (the same is responsible for the incompatibility of high-field expression Eq. (\ref {eq:highF}) of field-effects with the scaling). For the same reason consideration of an alternative\cite {caravaca10} to the energy balance equation Eq. (\ref{eq:hemSL}) will not help. Consequently, the onset field or the large-field conductance are not required to be power-laws in the HEM-SL. Note that even in the absence of scaling one can analyze the field/bias scale which may be obtained from Eq. (\ref {eq:hemSL}) as the following:
\begin{equation}
V_o \sim {(R_o T^{\beta})}^{1/2}.
\label{eq:hemfs} 
\end{equation}
There are however some special situations (see appendix B) where the HEM-SL does support scaling.

\section{Scaling in Two Dimension (2D)}
\subsection{Weak Localization (WL)} 
The hot electron model of Anderson et al.\cite {anderson79} allows complete determination of the field-dependent conductance as shown in the appendix A. Considering that the change in conductance in the WL regime is rather small,  Eq. (\ref{eq:lnRV}) can be recast as a power-law:
\begin{equation}
\Phi_{WL}  = {\left [1 + k^2 \left ( {V \over V_o} \right )^2 \right ]}^{z_F/2},
\label{eq:wlsc}
\end{equation}
where $z_F = \Sigma_u \lambda_V R_{\square o} ~ (\Sigma_u=e^2/ 2{\pi}^2 \hbar =1.2\times 10^{-5} ~ \Omega^{-1}$). $R_{\square o}=1/\Sigma_{\square o}$ is the temperature-independent sheet resistance. \textit{k} is a scale factor such that the field scale $V_o$ in Eq. (\ref{eq:wlsc})  matches the experimentally determined one. The later is defined such that $\Sigma(T,V_o)=(1+ \epsilon )~ \Sigma_o(T) $ where $\epsilon$ is an arbitrary chosen small number and $k^2 = {(1+\epsilon)}^ {2/z_F} - 1$. For $\epsilon \ll 1$, $k^2 \approx 2 \epsilon /z_F$. The field scale is then redefined (viz. Eq. \ref{eq:Afscale}) as
\begin{equation}
V_o = {\left ( {{|\epsilon| \gamma W} \over {a \Sigma_u \lambda_V}} \right )}^{1/2}~ T^{1+p/2}.
\label{eq:WLfscaleT}
\end{equation}
The same argument of small correction in Eq. (\ref{eq:2dRT}) allows one to write the Ohmic conductance as a power-law of temperature: $\Sigma_o \sim T^{\Sigma_u \lambda_T R_{\square o} }$. Using this in Eq. (\ref {eq:WLfscaleT}) yields
\begin{equation}
V_o \sim {\Sigma_o}^{x_T},
\label{eq:WLfscale}
\end{equation}
where
\begin{equation}
x_T = (1+p/2) / \Sigma_u \lambda_T R_{\square o} = 1/ \Sigma_u \lambda_VR_{\square o}  = 1/z_F,
\label{eq:wlxT}
\end{equation}
satisfying the SCR (Eq. \ref{eq:zT}). It follows from Eq. (\ref{eq:wlsc}) that at large bias the conductance varies as a temperature-independent power-law, $\Sigma(T,V) \sim V^{z_F}$. Thus, we have the theoretical field-dependent conductance in WL regime (Eqs.  \ref{eq:wlsc}-\ref {eq:WLfscale}) in complete agreement with the general scaling relations (Eqs. \ref{eq:scaling}-\ref{eq:fscale}).

\begin{figure}
\includegraphics[width=6.6cm]{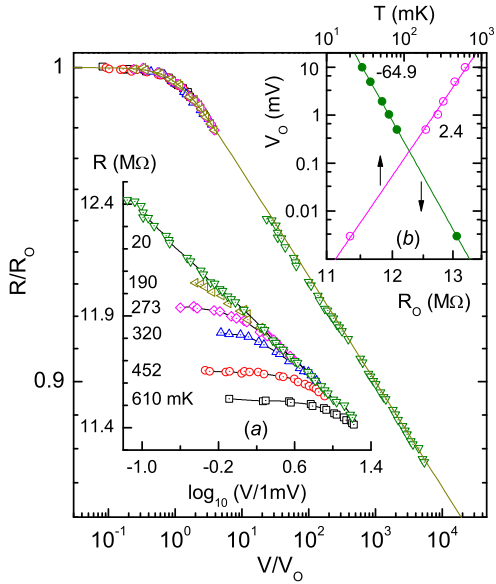}
\caption{(color online) Scaling $ (\epsilon \sim -0.005)$ in 2D weakly localized metallic thin films of AuPd. Plots of resistance vs. electric bias at different temperatures are shown in the inset (a) (data adapted from Ref. \onlinecite{dolan79}). Scaling of the same data leads to its collapse as shown in the main panel.  $R_o(T)$ and $V_o(T)$ are the Ohmic resistance and the onset bias respectively at temperature \textit{T}. The solid line is a fit to $\Phi_{WL} = {(1 + 0.9 q^2 )}^{1/130}~(q=V/V_o)$. Inset (b) shows log-log plots of $V_o$ vs. $R_o$ (solid symbols) and \textit{T} (open symbols). The solid lines are linear fits to the data with the slopes as mentioned. }
\label{fig.1}
\end{figure}
For comparison with experiment we revisit the data of Dolan and Osheroff\cite {dolan79} obtained in a 2D weakly localized thin film and shown in the inset (a) of Fig. 1. The data contain a lot more information than the simply logarithmic variation with the applied field, which has been the usual focus of such measurements so far. Indeed, the sample resistance \textit{R}(=1/$\Sigma$) with an initial Ohmic value ($R_o$) looks similar to those in strongly localized systems\cite {talukdar11}. There exists a voltage scale $V_o(T)$ at each \textit{T} such that $R \approx R_o$ for $q(=V/V_o) \leqslant 1$ and \textit{R} decreases for $q > 1$. As shown in Fig. 1 (main panel), the  collapse of all data at various temperatures on a single curve \cite {paper26_2} confirms scaling as indicated in Eq. (\ref{eq:wlsc}). The solid line in Fig. 1 indicates an excellent fit to the inverse of Eq. (\ref{eq:wlsc}) with $k^2 =0.9 ~\rm{for}~ \epsilon \sim -0.005$ and $z_F = 1/65$. The inset (b) displays a log-log plot of $V_o$ vs. $R_o$ which yields a large value\cite {paper26_3} for the nonlinearity exponent $x_T = 64.9 \pm 0.4$ confirming the SCR $z_F \approx 1/x_T$ (Eq. \ref{eq:zT}). With $R_{\square o}=4600 ~\Omega$ and $\lambda_V=0.28$ (Ref. \onlinecite {dolan79}), the theoretical value of $z_F$ is 1/62.9 which is close to the experimental one 1/65. One may note that Eq. (\ref{eq:WLfscaleT}) provides the most direct method for determining \textit{p}. A plot of $V_o$ vs. \textit{T} in the inset (b) yields $1+p/2 \approx 2.4$ close to 2.6 reported in Ref. (\onlinecite {dolan79}). Since $\lambda_V$ is of the order of unity and the maximum metallic sheet resistance\cite {imry} is predicted to be about $h/e^2 $ we have the minimum value $x^{min}_T \approx (\pi h /e^2) (e^2 /h) = \pi$.

To verify that $x_T \propto 1 / R_{\square}$, exponents (open symbols) processed from the data available in literature for five different materials (films) of thicknesses 3-13 nm and a 2DEG layer are shown in Fig. 2. All data are seen to lie on a line with a slope of -0.87 which is close to the expected value of -1. The small discrepancy may be due to the variation in experimental \textit{p}. Exponents with error bars were determined from the \textit{R-T} data that yield only $\lambda_T$ but not \textit{p} separately. We assumed \textit{p}=2 and $x_T$ was determined using Eq. (\ref{eq:wlxT}). Since the experimental range of \textit{p} is 1-3, error bars reflect the possible limits of $x_T$. The fitted line when extrapolated yields $x^{min}_T \sim 7$.
   
\begin{figure}
\begin{center}
\includegraphics[width=6.6cm]{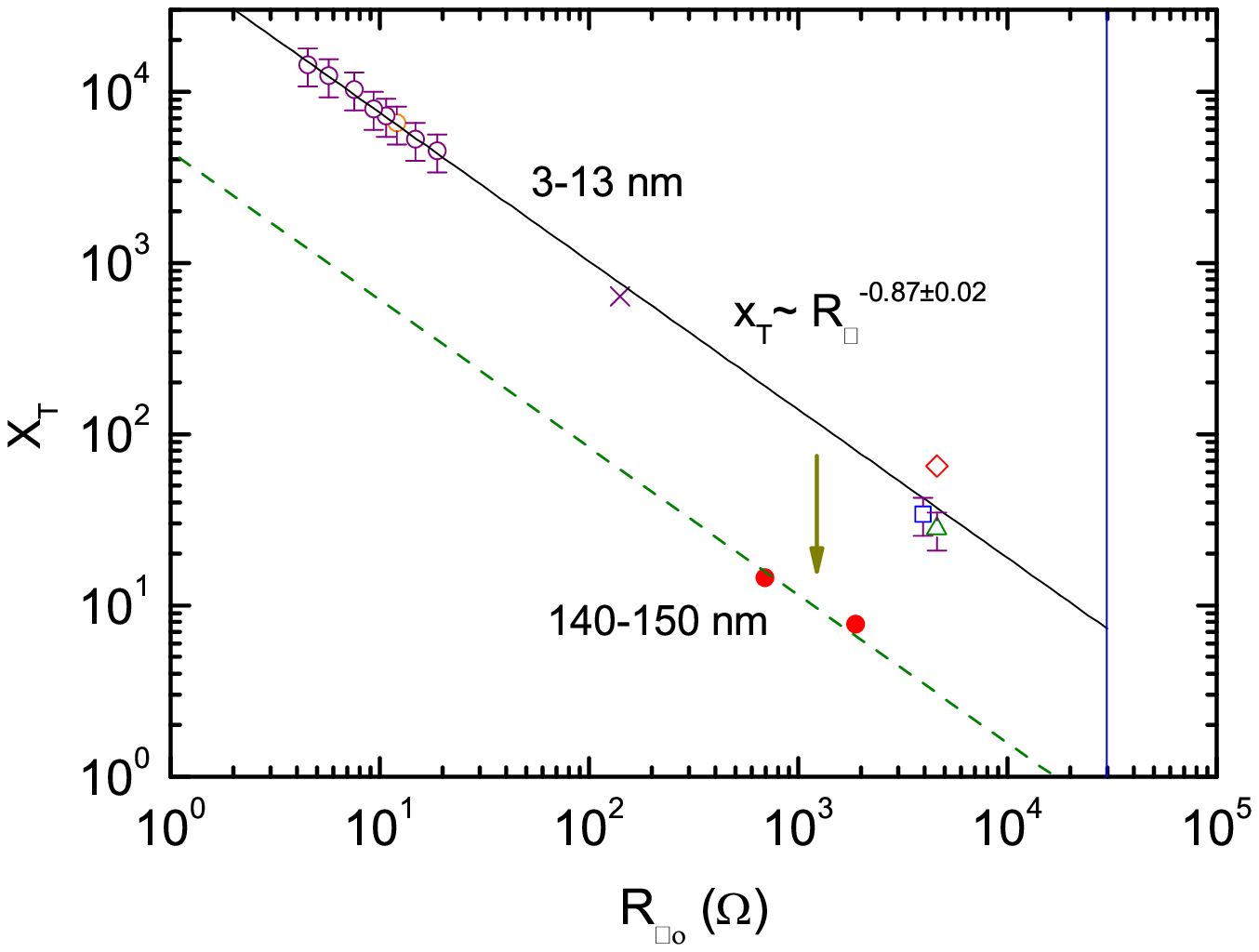}
\caption{(color online) Effect of film thickness on the nonlinearity exponent in 2D metallic states. The exponent is plotted against sheet resistance for various materials and film thicknesses. One set of data (open symbols) belongs to thin films ($\diamond$-AuPd [\onlinecite {dolan79}], $\square$-Pt[\onlinecite{hoffmann82}], $\triangle$-2DEG[ \onlinecite {gershen00}], $\times$-Au[\onlinecite {doroz82}]), and $\circ$-Cu[\onlinecite{[{}][{. $x_T$ corresponding to films of varying thicknesss ranged from 4500 (62$\AA$) to 14300 (130$\AA$).}]van81}]). See text for explanation on error bars. Another set of data (closed symbols) is from thick films (Refs. \onlinecite{osofsky88a}). The vertical arrow schematically indicates progress towards an incipient thickness-induced 2D-3D crossover. The solid line is a least-square fit and the dashed line is drawn parallel to it. The vertical line corresponds to the maximum metallic resistance of 30 k$\Omega$.}
\label{fig.2}
\end{center}
\end{figure}

Also shown in the figure are the data (closed circles) from two thick films (thickness 140-150 nm) of $\rm{Ge}_{0.79} \rm{Au}_{0.21}$ and $\rm{C}_x \rm{Cu}_{1-x}$ measured by Osofsky et al.\cite {osofsky88a,*osofsky88b}. Authors considered the films to be three dimensional and on the metallic side of the MIT, and analyzed the data using McMillan's scaling theory\cite {mcmillan81}. The later was interpreted to yield expressions for the Ohmic and field-dependent conductivity at small T's: $\sigma_o (T) = \sigma_o + \sigma_1 T^{1/2}$ and $\sigma (F) = \sigma_0 + \sigma_2 F^{1/3}$ respectively. Apart from the fact that $\sigma (F=0)$ is not equal to $\sigma_o (T)$ as it should be, salient features of \textit{T}- and \textit{F}-dependences are in doubtful conformity with the data. The reported small temperature variation of conductivities ($\sim T^{0.02-0.07}$) is akin to the logarithmic one characteristic of a 2D system. This suggests that these thick films may actually belong to the crossover regime from 2D to 3D. A film is considered to be 2D if the inelastic mean free path $l_{\phi}$ is greater than the thickness. Using $eF_o l_{\phi} \sim k T$ and $F_o \approx 9$ V/cm at 1.5 K, we estimate $l_{\phi} \sim 160 \rm{~nm}$ i.e. of the order of film thickness. We now show that the same model based on hot electron effects provides a much better and consistent description of the thick films as it did in thin films. Fig. 3 displays scaling of the conductivity data in the thick sample of $\rm{Ge}_{0.79} \rm{Au}_{0.21}$ at different temperatures. As seen the scaled data are well fitted (solid line) by Eq. (\ref{eq:wlsc}), thereby confirming that the thick sample is indeed in a 2D-3D transition regime. From the fit we get $z_F \approx 1/7.9$ and from the inset, $x_T = 7.8$. Thus the SCR is satisfied. This experimental value of $z_F$ is in clear contradiction with 1/3 predicted by the authors. Similarly we obtained $x_T$ = 14.5 for $\rm{C}_x \rm{Cu}_{1-x}$. Although the exponents from thick films separately follow the same trend with sheet resistance as thin films (Fig. 2), they are roughly an order of magnitude less than those in thin films of similar sheet resistances. This trend in the exponent values is in agreement with the experimental results in 3D strongly localized systems where $x_T$ is found to be less than 1 (see Tables). \textit{p} determined from the onset field data using Eq. (\ref {eq:WLfscaleT}) is 0.9 (see inset) and 0.7 for the two films respectively. These results indicate a possible renormalization of the constant $\alpha$ with the film thickness.

\begin{figure}
\includegraphics[width=5.7cm]{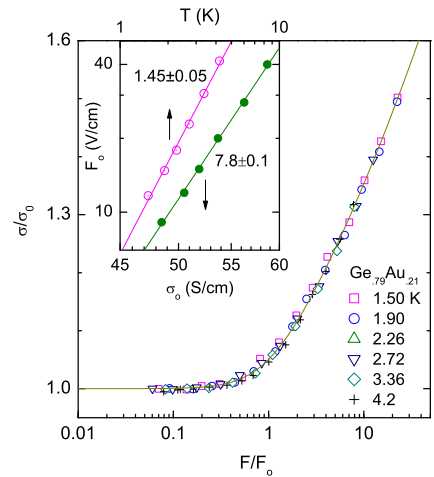}
\caption{(color online) Scaling $(\epsilon \sim 0.05)$ of the bias-dependent conductivity at different temperatures in a metallic thick film of GeAu (data adapted from Ref. \onlinecite{osofsky88b}). The solid line is a fit to $\Phi_{WL} = {(1 + 1.18 q^2 )}^{1/15.8}$. Inset shows log-log plots of $F_o$ vs. $\sigma_o$ (solid symbols) and \textit{T} (open symbols). The solid lines are linear fits to the data with the slopes as mentioned. }
\label{fig.3}
\end{figure}

Let us now consider the case when the control parameter is magnetic field. The megnetoresistance at small field\cite {lee85} is essentially determined by the ratio of two length scales corresponding to temperature and magnetic field: $L^2_{\phi} = D {\tau}_{in}$ and $L^2_H = \hbar c/4eH$. \textit{D} is the electron diffusion coefficient. Since $L^2_{\phi} \sim T^{-p}$ the replacing \textit{T} by $T_e$ as before gives the magnetoresistance as a function of a electric field. It is seen from Eq. (\ref {eq:Te}) that the bias scale should remain unaffected by the application of a magnetic field or,
\begin{equation}
x_B = 0.
\label{eq:xD}
\end{equation}
Physically, the zero exponent is justified since a magnetic field does not impart any energy to electrons.

Finally consider a situation where disorder is the control parameter, e.g. films with varying degree of disorder being measured at a fixed temperature. It follows from Eq. (\ref {eq:hem}) that the onset bias is given by $V_o \sim {\Sigma_o}^{-1/2}$, i.e.
\begin{equation}
x_D = -1/2.
\label{eq:xD}
\end{equation}
Since the exponent is negative the SCR is invalid and the nonlinear resistances are not required to be independent of disorder at large fields.

\subsection{Strong Localization (SL)}

As the conductance in 2D decreases from a large to a smaller value a crossover from WL to SL takes place as indicated by a change from logarithmic (Eq. \ref{eq:2dRT}) to activated (Eq. \ref{eq:mott}) temperature variation of the Ohmic conductance\cite {abrahams01}. We discuss here in detail the nonlinear conduction data of Gershenson et al. \cite{gershen00} in the SL regime of a 2DEG system. The sample exhibited a hopping transport with $T_o = 1.57$ K and \textit{m}=0.7. Authors claimed that the nonOhmic resistance in the SL regime is well described by the same equation (Eq. \ref{eq:mott}) for Ohmic resistance with \textit{T} replaced by an effective temperature $T_e$ (hot electron effects) given by Eq. (\ref{eq:hemSL}) with $C \propto e^2 R_{00} / h$, $R_{00} = 1/ \Sigma_{00} = 25 ~ k\Omega $ and $\beta = 4.5$. Unlike in WL, the electron-phonon interactions allow heat to flow out of the sample and determine the constant \textit{C}. A closer look at Fig. 5 of Ref. \onlinecite {gershen00} however reveals that fits which are indeed good at higher temperatures ($T \ge 0.215~K$) become progressively poorer at lower temperatures ($T \le 0.11~K$) (compare similar phenomena in 3D in Fig. 6 of Ref. \onlinecite {zhang98}). This correlates remarkably well with the scaling behavior at the two temperature ranges as shown in Fig. 5 (data at two higher temperatures left out for clarity). While the data in the higher temperature range (UTR) clearly do not collapse (as they should not according to the HEM-SL in Section IIA), those in the lower temperature range (LTR) do. Such difference in behavior manifests itself sharply when the bias scales are considered. According to Eq. (\ref{eq:hemfs}) the bias scale in the HEM-SL should vary as $V_o \sim { (R_o T^{\beta} )}^{1/2}$. A log-log plot of $V_o$ vs. $ R_o T^{4.5} $ in the inset (a) shows that while data in UTR do lie in a straight line those in LTR clearly deviate from the line. The slope of the fitted line is 0.59 compared to the expected value of 0.5. On the other hand, when $V_o$ is plotted against $R_o$ (inset b) a power-law  fit to the data in LTR yields an exponent $x_T \approx 0.08$ (Eq. \ref {eq:fscale}). It is seen that the exponent $x_T$ is greater than 1 in the Wl regimes (of previous examples) while it (i.e., 0.08) is less than 1 in the SL regime.

\begin{figure}
\includegraphics[width=6.5cm]{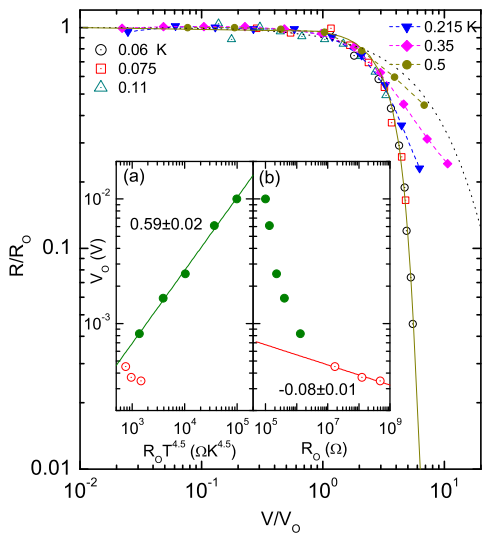}
\caption{(color online) Evidence for a temperature-induced transition in the SL regime. In a 2DES sample, the bias-dependent conductivity data scale only at lower temperatures $(\epsilon \sim -0.1) $. The data are adapted from Ref. \onlinecite {gershen00}. The solid line is a fit to $\Phi_H^{0.16} - 1/ \Phi_H = {(0.227 q)}^2$ and the dotted line to $\Phi = \exp (-0.105q)$. Insets show log-log plots of $V_o$ vs. $ R_o T^{4.5}$ (a) and $R_o$ (b). The solid lines are least-square fits to the selected data with the slope as shown. }
\label{fig.4}
\end{figure}

\begin{figure}
\includegraphics[width=6.5cm]{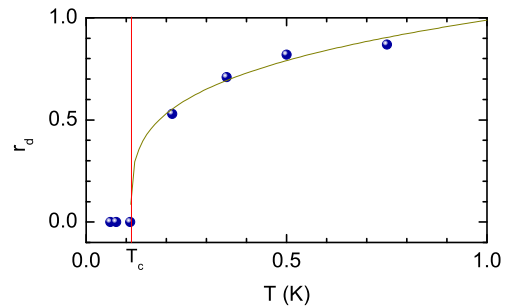}
\caption{(color online) $r_d$ vs. \textit{T} for the scaling-nonscaling transition (SNST) in Fig. 4. The curve is a fit to $r_d = 1.02 {(T - 0.111)}^ {0.27}$. }
\label{fig.5}
\end{figure}

Clearly, \textit{within} the SL regime there is a transition from one mechanism of conduction obeying the scaling at low \textit{T} to another governed by hot electron effects at higher \textit{T} at about $T_c \sim 0.11$ K ($T_o/T \sim 14$) (inset a). In the appendix B, a phenomenological picture is considered where hot electron effects remain valid on both sides of this scaling-nonscaling transition (SNST) with an additional, and yet unknown, mechanism that gives rise to the scaling along with Eq. (\ref{eq:fscale}). The picture predicts a transcendental equation $\Phi_H^{2x_T} - \Phi_H^{-1} = {(kq)}^2$ (Eq. \ref{eq:hemsf}) which is seen to fit (solid line) the data reasonably well with the same $x_T=0.08$ albeit deviating slightly near the 'bend' near $q \sim 1$. The scale factor \textit{k} is given by $k^2 = {(1+ \epsilon)}^{2 x_T} - {(1+ \epsilon)}^{-1}$. The fit without any free parameter is yet an another affirmation of the validity of the scaling proposal. The fit indicates that the large field exponent $z_F$ is about $ 1/0.08 \approx 1/x_T$ so that the SCR (Eq. \ref{eq:zT}) is satisfied. This transition appears continuous and is temperature-induced in contrast to the WL-SL crossover which is disorder-induced (i.e., $\xi$ varying). A similar transition ($T_o/T \sim 135$) near a metal-insulator transition in 3D was tentatively indicated by Zhang et al.\cite {zhang98} (discussed further in Section IV.A). The transition can be described quantitatively by monitoring the ratio $r_d (T) = {(\Sigma_o / \Sigma)}_d = {(R / R_o)}_d$ as a function of temperature where ${(R / R_o)}_d$ is the value of the normalized resistance at which the normalized resistance curve at $T \geq T_c$ deviates from the scaled curve. As \textit{T} is reduced from a high value to $T_c$, $r_d$ is expected to vary continuously from about 1 to 0 at $T_c$ as seen in Fig. 5. The curve fitted to a power-law is quite suggestive but must be considered tentative given insufficient data near the transition and uncertainties in $r_d$ at higher temperatures.

\section{Scaling in Three Dimension (3D)}
As disorder increases, unlike a crossover in 2D, an actual transition from a metallic state to an insulating state takes place in 3D. Like in a standard critical phenomenon, the localization length $\xi$ diverges at the transition upon approaching from the insulating side. $\xi$ decreases upon increasing disorder as one moves away from the transition further into the insulating regime. Both large and small localization length regimes are strongly localized ones where, as mentioned earlier, the conduction is activated and characterized by a temperature variation such as given by Eq. (\ref {eq:mott}). We apply below the scaling formalism to both the regimes and highlight model-independent features of the effects of a field in 3D and challenges faced by inadequate theories.

\subsection{Metal-Insulator Transition (MIT) (large $\xi$)} 
A regime in the vicinity of a MIT with large $\xi$ is similar to the 2D case just discussed and is usually studied in doped semiconductors with a critical doping concentration $n_c$. The diverging localization length is given by $\xi \sim  {\delta_n}^{-\nu}$ where $\delta_n = |(n-n_c)|/n_c \ll 1$. Phenomenologically, the conductivity in the critical region at a fixed temperature \textit{T} is given by a power-law: $\sigma_o(n) \sim { \delta_n }^{\mu}$ where $\nu$ and $\mu$ are exponents. Depending upon the system length scale \textit{L} we discuss below two situations.

$\xi > L$ (critical regime): At a fixed doping \textit{n} the conductivity is given by $\sigma_o \sim e^2 / \hbar L_M$\cite {imry} where $L_M$ is an appropriate length scale. The physics here is determined by two length scales\cite {larkin82,sondhi97} corresponding to temperature and electric field: $L_{\phi} \sim T^{-1/\eta}$ and $L_F \sim F^{-1/(1+\eta)}$ with $1 < \eta < 3$. The exponent $\eta$ relates the energy scale with the length scale. This means that for small field $\sigma_o \sim T^{1/\eta}$ and at large field $\sigma \sim F^{1/(1+\eta)}$ so that $z_F = 1/(1+\eta)$. The onset field is determined when $L_{\phi} \sim L_F$. This yields $F_o \sim T^{(1+\eta)/\eta} \sim \sigma_o^{1+\eta}$ so that $x_T = 1 + \eta $. Hence the SCR (Eq. \ref {eq:zT}) is satisfied. Note that $x_T > 2$  as $\eta > 1$.

$\xi < L$: There are many studies on the insulating side of the MIT. Some of those which exhibit the scaling property are shown in Table I (under doped semiconductors). There are others\cite {wang90,zhang98,galea07} which are amenable to the HEM-SL do not. There is however no understanding at present as to why one system does follow the scaling and another does not. We show that even those systems in 3D described by HEM-SL exhibit the same transition to scaling regimes at lower temperatures as in 2D. We consider two sets of data from two different samples, \textit{a} (Zhang et al.\cite {zhang98}) and \textit{b} (Galeazzi et al.\cite {galea07}, $\delta \sim 9/100$), of the same system of doped Si:P:B, 50\% compensated, taken at two marginally overlapping ranges of temperatures. The first set of data \textit{a} were taken at 0.393-0.1 K and the second set of data \textit{b} at 0.16-0.06 K. In both cases, \textit{m}=1/2 and $\beta \approx 5.5$. $T_o$ was 4.73 and 8.55 K respectively. It is seen in Fig. 6 that data \textit{a} at different temperatures do not collapse on a single curve (as they should not in the HEM-SL). On the other hand, data \textit{b} at lower temperatures ($T \leq 0.09$ K or $T_o/T \geq 11$) nicely collapse on to a curve while the same at higher temperatures $\geq 0.1$ K like data \textit{a} do not (not shown), indicating a transition as in 2D. The same is again indicated by the log-log plots of  $V_o$ vs. $ R_o T_o^{5.5} $ in the inset. $V_o$ for sample \textit{a} (closed symbols) fall on a line with a slope of 0.56 compared to the expected 0.5 (Eq. \ref{eq:hemfs}). The bias scale in sample \textit{b} (open symbols) turns out to be nearly a constant within error (i.e., $x_T \approx 0$) and is off the straight line (compare with the inset \textit{a} of Fig. 4. Note that the exact continuity in the data belonging to two different samples can not be expected). Using the same picture as in 2D, data were fitted to $\Phi_H^{-1} = 1 - {(kq)}^2$ (Eq. \ref {eq:hemsf0}) corresponding to $x_T=0$. \textit{k} is the usual scale factor given by $k^2 = 1 - {(1+ \epsilon)}^{-1}$. As seen in Fig. 6 the fit (solid line) is excellant, better than the one in Fig. 4. This point is furthur discussed in the appendix B.

\begin{figure}
\includegraphics[width=6.4cm]{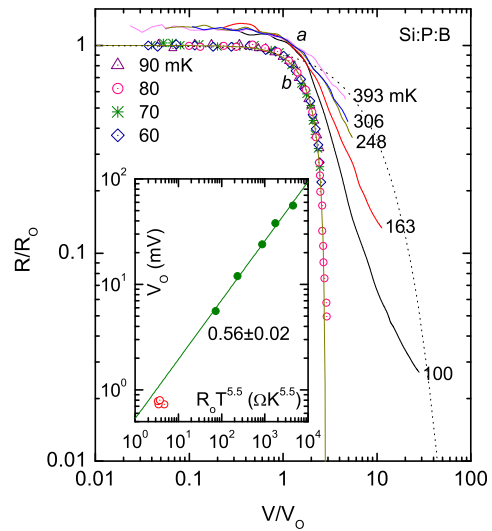}
\caption{(color online) Scaling $ (\epsilon \sim -0.1)$ of the bias-dependent conduction data at different temperatures in a \textit{n}-type Si:P:B sample on the insulating side of the MIT. The data \textit{a} (shifted upwards by a factor 1.2 for clarity) and \textit{b} are adapted from Ref. \onlinecite {zhang98} and \onlinecite {galea07} respectively. The solid line is a fit to $\Phi_H^{-1} = 1 - {(0.354 q)}^2$ and the dotted line to $\Phi = \exp (-0.105q)$. Inset shows log-log plots of $V_o$ vs. $R_o T^{5.5}$ corresponding to data \textit{a} (closed symbols) and data \textit{b} (open symbols). The solid line is a least-square fit to the data (closed symbols) with the slope as shown. }
\label{fig.6}
\end{figure}

\begin{figure}
\includegraphics[width=6.5cm]{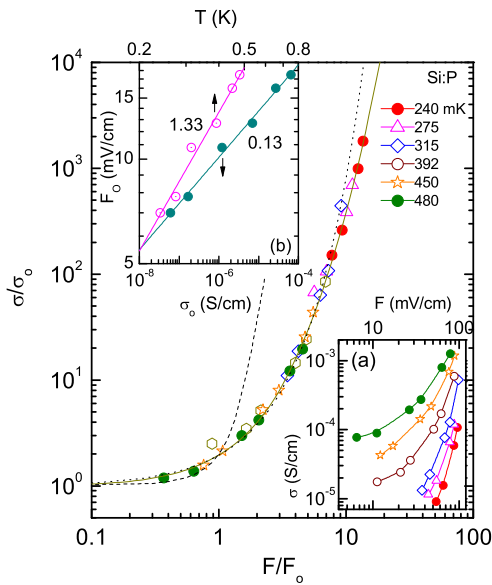}
\caption{(color online) Scaling $ (\epsilon \sim 1)$ of the bias-dependent conductivity data at different temperatures in a doped Si:P sample on the insulating side of the MIT. The data are adapted from Ref. \onlinecite {rosen80}. The solid line is a fit to the SGM expression\cite {talukdar11}: $\Phi = 1+0.9q^{4/3}+ 0.06 q^{18/5} +2.2 \times 10^{-5}q^{27/4}$, dashed line to $\Phi_H^{0.3} - 1/ \Phi_H = {(0.855 q)}^2$ and the dotted line to $\Phi = \exp (0.67q)$. Inset shows two log-log plots of $F_o$ vs. $\sigma_o$ (closed symbols) and \textit{T} (open symbols). The solid lines are least-square fits to the data with the slopes as shown. }
\label{fig.7}
\end{figure}

Next we consider a system similar to the one above but not amenable to the HEM-SL - Si:P samples (Rosenbaum et al.\cite {rosen80}) on the insulating side of the MIT ($\delta \sim 1/100$). The behavior of $\sigma_o (T)$ was rather ambiguous in that the data apparently could be fitted to both power-law and exponential with either \textit{m}=1/4 or 1/2 apparently without any noticeable differences in fitting qualities. The power-law fit yields an unphysically small $1/ \eta \approx 1/10$ (theoretical value is about 1) and hence, discarded in favor of the exponential with \textit{m}=1/4 and $T_o \approx 10^6~K$. The reason for choosing 1/4 over 1/2 is given below. The inset (a) in Fig. 7 displays conductivity data at various temperatures, which are scaled in the main panel. The quality of the data collapse confirms the applicability of the scaling. In fact, the fit to SGM (solid line) suggests that the large-field conductivity is a power-law with the exponent $z_F \approx 6.75$. The onset fields $F_o$ obtained in the process of scaling are plotted against both $\sigma_o$ and \textit{T} in the inset (b) with slopes $x_T = 0.13$ and 1.33 respectively. According to Eq. (\ref {eq:lowF}), $F_o \sim T^{1+m\mu}$ with $\mu$=1 or 2\cite {talukdar11}. Thus, \textit{m}=1/4 is in better agreement with the experimental value of 1.33. $z_F$ is close to 1/$x_T=7.6$ so that the SCR (Eq. \ref{eq:zT}) is satisfied. Compared to $x_T>1$ in WL regimes and the critical region of the MIT, $x_T$ is less than 1 in the SL regimes. This is in line with the values generally found in strong localized systems (see Tables). Fig. 7 also shows a plot (dashed line) to Eq. (\ref{eq:hemsf}) passing through the data at \textit{q}=1 and then rising much faster. On the other hand, the plot (dotted line) to Eq. (\ref{eq:lowF}) is fairly good up to about $\sigma / \sigma_o \sim 80$. This gross mismatch between a scaling state in absence of any scaling-nonscaling transition and one in a transition clearly indicates that a physical mechanism akin more to field-effects than hot electron effects is at play at least in the present sample.

\subsection{Strong Localization (small $\xi$)}
Let us now move to systems with strong disorder and localization length much smaller than ones considered so far. In reality, there are many such classes of materials. We refer to Refs. \onlinecite {talukdar11} and  \onlinecite {talukdar12} for recent scaling analysis of field-dependent data in the two classes of materials as far apart as conducting polymers and manganites respectively. The nonlinearity exponents obtained in these materials are shown for convenience in Tables II and III in Appendix C. We have subjected the scaling analysis to several other classes of materials including classical systems such as amorphous and doped semiconductors whose data are available in literature. The exponents along with other relevant information obtained in these materials are displayed in Table I. All the samples in Table I except the organic crystals exhibit VRH conduction. As a further example of model-independent scaling, let us discuss this non-VRH, correlated system, namely layered organic crystals $\theta$-(BEDT-TTF)$_2$CsZn(SCN)$_4$ by examining the experimental data and the proposed model\cite {takahide06}.

\begin{figure}
\includegraphics[width=6.5cm]{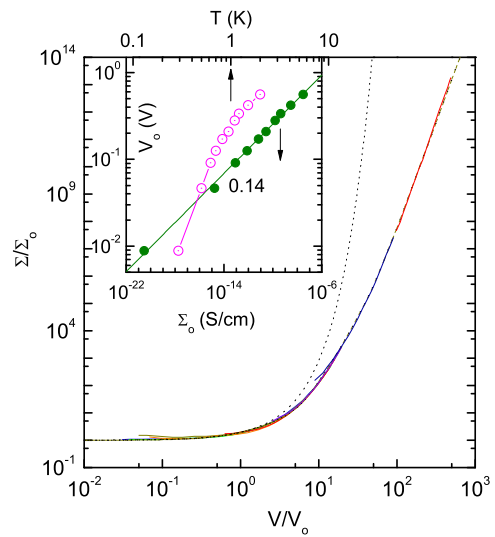}
\caption{(color online) Scaling $(\epsilon \sim 1)$ of the bias-dependent conductance data in a correlated, charge-ordered organic salt at ten different temperatures (0.29-2 K) (data from Ref. \onlinecite {takahide06}). The line mode for data representation is used to highlight the quality of data collapse. The solid line is a fit to the SGM expression\cite {talukdar11}: $\Phi = 1+0.97q^{4/3}+ 0.0025 q^{14/3} +1.5 \times 10^{-8}q^{70/9}$ and the dotted line to $\Phi = \exp (0.67q)$. Inset shows two log-log plots of $V_o$ vs. $\Sigma_o$ (closed symbols) and \textit{T} (open symbols). The solid line is a linear fit with the slope as shown. }
\label{fig.8}
\end{figure}

Disorder arises out of random presence of holes in the charge-ordered layers. The low field transport in the insulating phase of these crystals is characterized by an activated process with \textit{m}=1, $T_o$=24 K. The \textit{I-V} characteristics were measured at different temperatures and found to follow the power-law at large bias with a large exponent ($\sim 8.4$) similar to observations in many systems including conducting polymers\cite {talukdar11}, \textit{a}-Ge \cite {morgan71} or \textit{a}-Si:H\cite {nebel92}. Authors suggested a model based on unbinding of electron-hole pairs by thermal excitation in the background of the charge-ordered ground state. The power-law characteristics was given by $I \sim V^{1~+U_o/2kT}$  so that $z_F = U_o/2kT$. Furthermore, the onset (or, crossover) field was given by the temperature-independent $F_o = U_o/e \lambda$ implying $x_T = 0$. Here $U_o$ is the strength of the Coulomb potential and $\lambda$ is the screening length. Thus, according to the model $x_T \neq 1/z_F$. Experimental data normalized according to Eq. (\ref{eq:scaling}) are presented in Fig. 8. The high quality of the collapse of the data at various temperatures points toward validity of the scaling in this correlated system. Ohmic conductances at lower \textit{T} were obtained from extrapolation of the \textit{$\Sigma_o$-T} data. The inset shows plots of $V_o$ vs. $\Sigma_o$ and \textit{T}. The straight line through the closed circles proves the validity of Eq. (\ref{eq:fscale}) (over a range of 14 decades!) and yields a non-zero exponent $x_T =0.14 \pm 0.01$ in contradiction to the model prediction of zero. From the fit (solid line) to SGM expression we assume $z_F \sim 7.8$ close to 7.4 given by the authors. Therefore, $z_F \sim 1/x_T$ satisfying the SCR (Eq. \ref{eq:zT}). $\lambda$ was determined using $F_o$ which varies by about two orders of magnitude (inset) within the experimental range. Therefore, this casts doubt on the conclusion (i.e., the long-ranged Coulomb interaction) drawn from a particular value of $F_o$. Furthermore, the model expression of a temperature-dependent $z_F$ is untenable, particularly at low \textit{T} and at high fields when the conductivity becomes temperature-independent i.e., 'activationless' as reported by the authors.

\section{Discussion and Conclusions} 
The principal aim of this paper was to establish validity of the scaling relations Eqs. \ref{eq:scaling}-\ref{eq:fscale} as the universal ones capable of describing the field-dependent data across the whole spectrum of disorder. Various examples in different disorder regimes discussed in Sections III and IV validate the hypothesis except in certain cases which are better described by hot electron effects. The scaling in WL regimes is now well understood both theoretically and experimentally but the scaling in SL regimes in both 2D and 3D is rather phenomenological although it has been possible to predict the scaling function in some cases. Some heuristic arguments (Section II) may provide some rationale particularly for the novel scaling of the field scale (Eq. \ref{eq:fscale}) but the scaling in SL regimes poses a major theoretical challenge and adds to the list of difficult problems of transport critical phenomena in disordered systems. The strongly disordered systems in 3D comprise of an wide variety of systems - amorphous/doped semiconductors, conducting polymers, organic crystals, manganites, composites, metallic alloys, double perovskites (see Tables) - ranging from strongly correlated ones to weakly correlated ones. These diverse systems are all well described by the universal scaling\cite {unpub26_1} (see next paragraph). Yet, there exist presently separate theories for separate systems, often falling short of fulfilling the general requirements of the model-independent scaling as seen, for example, in cases of organic crystals or VRH systems. This situation is really reminiscent of the pre-Landau era in thermodynamic critical phenomena. A theoretical framework which would be independent of microscopic details as in 2D would be highly desirable\cite {staveren91}. Indeed, as discussed below, lack of such a theory prevents full understanding of the unexpected, nonuniversal structure of the nonlinearity exponents.

\subsection{Scaling Functions and Types of Strongly Localized Materials}
A SL regime is easily distinguished from a WL regime by the huge scale of change in conductivity brought about by the application of a field. As to be expected, there are variety of scaling functions each of which is suited for a particular regime of disorder. The exact function for the WL regimes is given by Eq. \ref {eq:wlsc}. The scaling function with only one adjustable parameter (i.e., high-field exponent) generates excellent fits to the experimental data (Figs. 1 and 3) and is in full conformity with the universal scaling. An approximate function (Eq. \ref {eq:hemsf}) containing the relevant nonlinearity exponent as the only parameter is suggested for the scaling state of the SNST in the SL regimes in both 2D and 3D. This function also generates good fits to the experimental data (Figs. 4 and 6). The fact that the exponent remains theoretically unexplained illustrates the challenges in the SL regimes. Most systems in 3D SL regimes do not undergo the SNST and the scaling functions in such cases have been conveniently and consistently fitted with the scaled version of Glazman-Matveev (SGM) expression\cite {talukdar11} although its justifiability at this moment may be a moot issue. There are four SL examples discussed in this work. $x_T$ has a value 0.08 in one (Fig. 4), 0 in another (Fig. 6) and about 0.14 in two others (Figs. 7 and 8). The corresponding scaled curves convey a reasonable impression of diversity in the scaling functions involved in the process keeping in mind that $\epsilon$ is -0.1 for the first sample, and 1 for the rest. Note in particular that the two scaling functions corresponding to nearly same exponent $\sim 0.14$ are not quite same. For comparison, fits (dotted lines) to the 'field-effect' expression $\exp (kq)$ (Eq. \ref {eq:lowF}) are also shown in the four figures. The scale factor \textit{k} was chosen such that $\exp (k)= 1+\epsilon $ i.e. each fit was made to pass through the data points at \textit{q}=1. The exponential is wide off the data for $q>1$, rising too slowly in Figs. 4 and 6. This is according to the fact that hot electron effects dominate these systems. In comparison, the fit in n-Si:P (Fig. 7) is quite good up to about $\sigma / \sigma_o \sim 80$ and beyond that (i.e., at low temperatures, high fields), rises faster than the data indicating perhaps dominance of field-effects over hot electron effects. Surprisingly, the same kind of agreement is not observed in the charge-ordered salts (Fig. 8) where the exponential rises too fast. 
  
It is observed that the scaling behavior can be used to categorize the strongly localized materials into three types: Type I, comprising of materials which obey the scaling according to Eq. \ref{eq:scaling} in full range of applied fields and have the high field conductances given by power-laws (Eq. \ref {eq:zT}); Type II, comprising of materials which obey the scaling as above but have the high field conductances given by exponentials (Eq. \ref{eq:highF}) and branching off the scaled curve. It is further found that \textit{m} for these materials is invariably 1/2. Further details are planned to be published elsewhere; Type III, comprising of materials which do not obey the scaling according to Eq. \ref{eq:scaling}. Types are indicated in Tables I and II but not in Table III since the high field behavior of manganites remains unclear.

\subsection{Scaling-Nonscaling Transition (SNST)}
The scaling analysis led to the interesting discovery of a temperature-induced scaling-nonscaling (S-NS) crossover in both 2D and 3D SL regimes. The crossover appears continuous and takes place at $T_o/T \sim 14$ in a 2DEG (Section III.B) and at about 11 (Section IV.B) and 135 (Ref. \onlinecite {zhang98}) in 3D samples near the MIT (same system but with different dopant amount). Thus, the ratio does not appear to have an universal value. The phase at higher temperatures is described by the HEM-SL. The picture in appendix B assumes that the HEM-SL continues to apply to lower temperatures but with an added mechanism that enforces Eq. \ref{eq:fscale}. More work is needed to determine what other properties characterize the two phases. The transition is however somewhat counterintuitive in that one would expect the 'pure' HEM-SL to apply at lower temperatures since the thermal relaxation would be very sluggish resulting in an electron-phonon bottleneck. But in reality, the opposite is true. There has been a suggestion\cite {marnieros00} of a crossover from electron-electron interaction-assisted hopping at low \textit{T} to phonon-mediated hopping at high \textit{T}. It is not clear how this crossover could be related to the present SNST. In addition, one has to reckon with the fact that not many disordered systems (see Tables) comply with the HEM-SL. One possibility could be that $T_c$'s in those systems are higher than the temperatures employed in the experiments. However it must be rejected considering that highest temperatures employed usually lay in the range of already high temperatures, 77-300 K. Real reasons will possibly be clear once we have better understanding of the physical conditions that trigger the onset of the scaling. The hot electron model does not depend on the microscopic details or dimensionality. Given its generality, it is surprising that it applies only to systems near the MIT in 3D.

\subsection{Nonlinearity exponents}                 
\begin{figure}
\includegraphics[width=6cm]{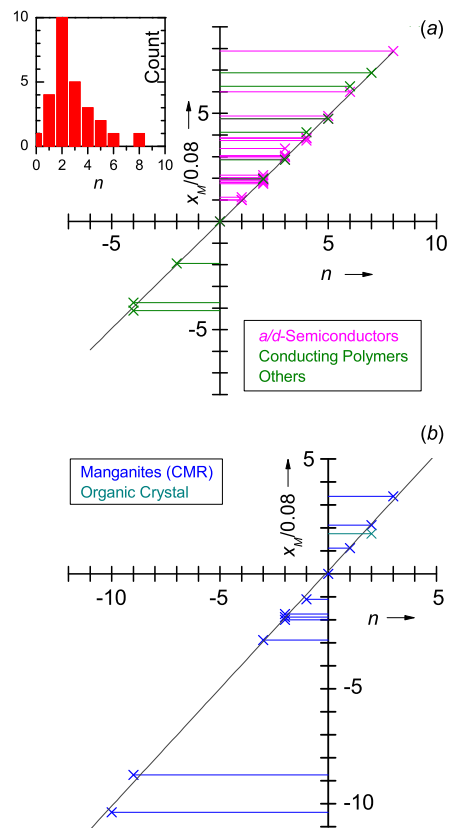}
\caption{(color online) The nonlinearity exponents $x_T$ in several (\textit{a}) uncorrelated and (\textit{b}) correlated classes of materials. The exponents are quantized and given by $x_M \approx 0.08n$. This is highlighted by plotting $x_M /0.08$ against \textit{n} with lines dropped off the data points to y-axes. Each panel shows a line passing through the origin with a slope of unity. Data points either lie on, or are very close to, the lines. The inset in panel (\textit{a}) displays the histogram of the exponent $x_T$ in Table I only.}
\label{fig.9}
\end{figure}
                 
\begin{figure}
\includegraphics[width=6cm]{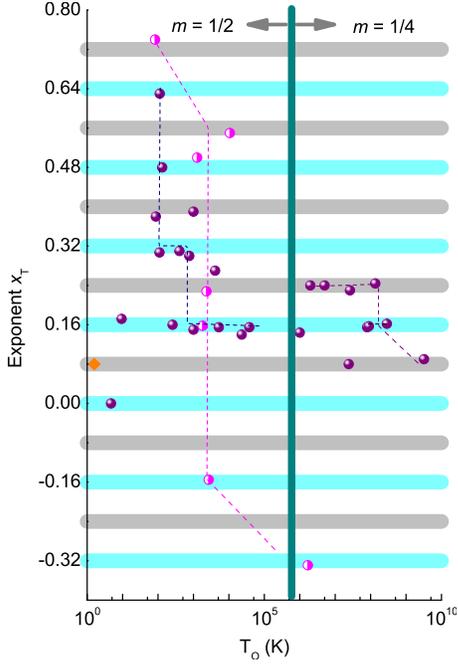}
\caption{(color online) Plot of $x_T$ vs. $T_o$ in VRH systems (Tables I and II). Horizontal bands, each 0.02 wide, are drawn at $x_T = 0.08n$ where \textit{n} is an integer. Data from Table I (purple closed symbols) including one in 2D (diamond) and Table II (magenta half-open symbols) are included. The vertical line roughly separates points according to its \textit{m} values. Dashed lines are only guides to the eye. Most of the points are seen to lie within the bars.}
\label{fig.10}
\end{figure}

Recall that $x_T \propto 1 / R_{\square}$ in WL. Such a material-dependent critical exponent presents a phenomenon that is far removed from the concept of universality in standard critical theory. It has been discussed in length in Section III and fully explained. Also it has been seen how this exponent helped in monitoring of an incipient dimensionality crossover as the film thickness increases (Fig. 2). Among the SL examples discussed above one finds diversity in the values of the exponent $x_T$: 0.08 in 2D and 0, 0.13, 0.14 in 3D. Exponents in many other systems are displayed in Table I. Exponents in conducting polymers and manganites already reported in Refs. \onlinecite {talukdar11,talukdar12} are shown for convenience in Tables II and III. It is apparent that not only there is diversity in exponent values across systems there is also diversity \textit{within} a given system. There are three such examples in the tables - \textit{a}-Ge and \textit{n}-Si:As in Table I and doped PPy in Table II. The five samples of \textit{a}-Ge from different laboratories yielded two values of 0.16 and three values of $\sim$0.24 for $x_T$. The five samples of doped polypyrrole yielded as many values with two of them being negative (-0.16 and -0.33). Two Si:As samples were prepared by the same laboratory using the same method. Obviously, one must look beyond the concept of universality in treating these results. On a closer inspection of the exponents a remarkable fact emerges. An exponent, positive or negative, could be described as an integer multiple of a number $\approx 0.08$ i.e., $x_T \approx 0.08 n $ where \textit{n} is an integer which can take both positive and negative values including zero. The number, 0.08 is thus the largest common factor among the exponents in the tables. To highlight the quantized nature of the exponents, $x_M / 0.08$ is plotted against \textit{n} for weakly- or non-correlated materials in Fig. 9a, and for correlated materials in Fig. 9b with lines dropped off the data symbols to \textit{y}-axes. The solid line in each figure has a slope of unity and passes through the origin. Points are seen to be lying close to the lines and distinct bands form around integer values on the \textit{y}-axis, confirming the quantized behavior. The width and density of a band reflect the spread in the values, and the frequency, respectively of the associated exponent found so far. What is truly impressive and striking is the fact that the exponents follow the same relation in systems such as amorphous semiconductors and manganites which are otherwise so physically distinct. Existence of such a common property in correlated and uncorrelated systems is indicative of some universal physics at play in the field-dependent conduction in strongly disordered systems.

To find out any possible correlation between the exponent $x_T$ and the corresponding characteristic temperature $T_o$ (Eq. \ref {eq:mott}) in VRH systems, these quantities for $m < 1$ in Tables I and II are plotted as shown in Fig. 9. Points generally lie within, or on the borders of, the horizontal bands of width of 0.02. There are hints of plateaus although presence of considerable scatter (each point representing a different sample) particularly for \textit{m}=1/2 prevents reaching definite conclusions. However, at least in 3D a broad trend of $x_T$ decreasing with $T_o$ is quite discernible for both \textit{m}=1/2 and 1/4. With $T_o \sim {\xi}^{-1}~\rm{and}~ {\xi}^{-3/4}$ respectively, it follows that $x_T$ decreases as $\xi$ decreases or disorder increases. As mentioned earlier this is in line with the general trend across all regimes of disorder although the rate of variation, albeit discreet, is much smaller in the SL regime. The plot is derived mostly from three classes of materials - conducting polymers, amorphous semiconductors, and doped crystalline semiconductors and reveals interesting differences among these classes. The semiconductors exhibit only positive exponent whereas conducting polymers (as well as manganites) exhibit both positive and negative exponents. It may be noted that the field-effect models do not admit any the negative exponent for $x_T$\cite {talukdar11}. Implications of the sign on the $\sigma - F$ curves have been illustrated in several figures of Ref. \onlinecite {talukdar11}. Furthermore, compared to doped semiconductors the highest quantization number \textit{n} for amorphous semiconductors (\textit{m}=1/4) analyzed so far seems to be 3. It remains a moot issue whether each class of materials follows its own separate trajectory as indicated for conducting polymers.

There are couple of experimental values available for $x_D$ (corresponding to disorder as a control parameter) in 3D - 0 in \textit{a}-Si:H (Table I) and -0.3 in doped PPy (Table II). The hot electron model yields a negative exponent, $x_D = -0.5$ (Eq \ref{eq:xD}) in 2D WL regimes. In fact, the prediction is irrespective of dimensionality. Thus, at least the sign of the exponent in 3D systems agrees with that in the HEM which however applies to a very limited number of 3D systems. Interestingly, field-effect models appear to agree qualitatively with the trend implied by a negative $x_D$ in that the onset field decreases as the Ohmic conductivity increases. Considering for simplicity the specific case of \textit{m}=1/2 and a fixed temperature we have from Eq (\ref{eq:lowF}) $F_o \sim L^{-1}_h \sim {\xi}^{-1/2}$ since $L_h \sim R_h \sim \xi {T_o}^{1/2} \sim \xi {\xi}^{-1/2}$. Since $\sigma_o$ is an increasing function of $\xi$, $F_o$ will decrease if $\sigma_o$ increases. However quantitative compliance is doubtful. The data in different samples of Ni$_x$SiO$_{2~1-x}$ (Fig. 28 of Ref. \onlinecite{abeles75}) indicate a negative $x_D$. An another value of -0.16 for the same exponent was obtained in ZnO-based varistors\cite {unpub26_2}. All these evidences strongly suggest that at least in 3D $x_D$ like $x_T$ also follows the same quantized relation, $x_D \approx 0.08 n $ but for $n \leq 0$. A strongly disordered system is generally modeled after a percolating system\cite {shklovskiibook}. However above the percolation threshold, and at room temperature, $x_D$ is found to be positive both experimentally\cite {cbb91,gefen86} and theoretically \cite {gefen86}. It will be interesting to determine $x_D$ on the insulating side of the MIT by varying the dopant concentration. Note that the 2D model prediction of -0.5 is not an integer multiple of 0.08.
                 
\begin{figure}
\includegraphics[width=6cm]{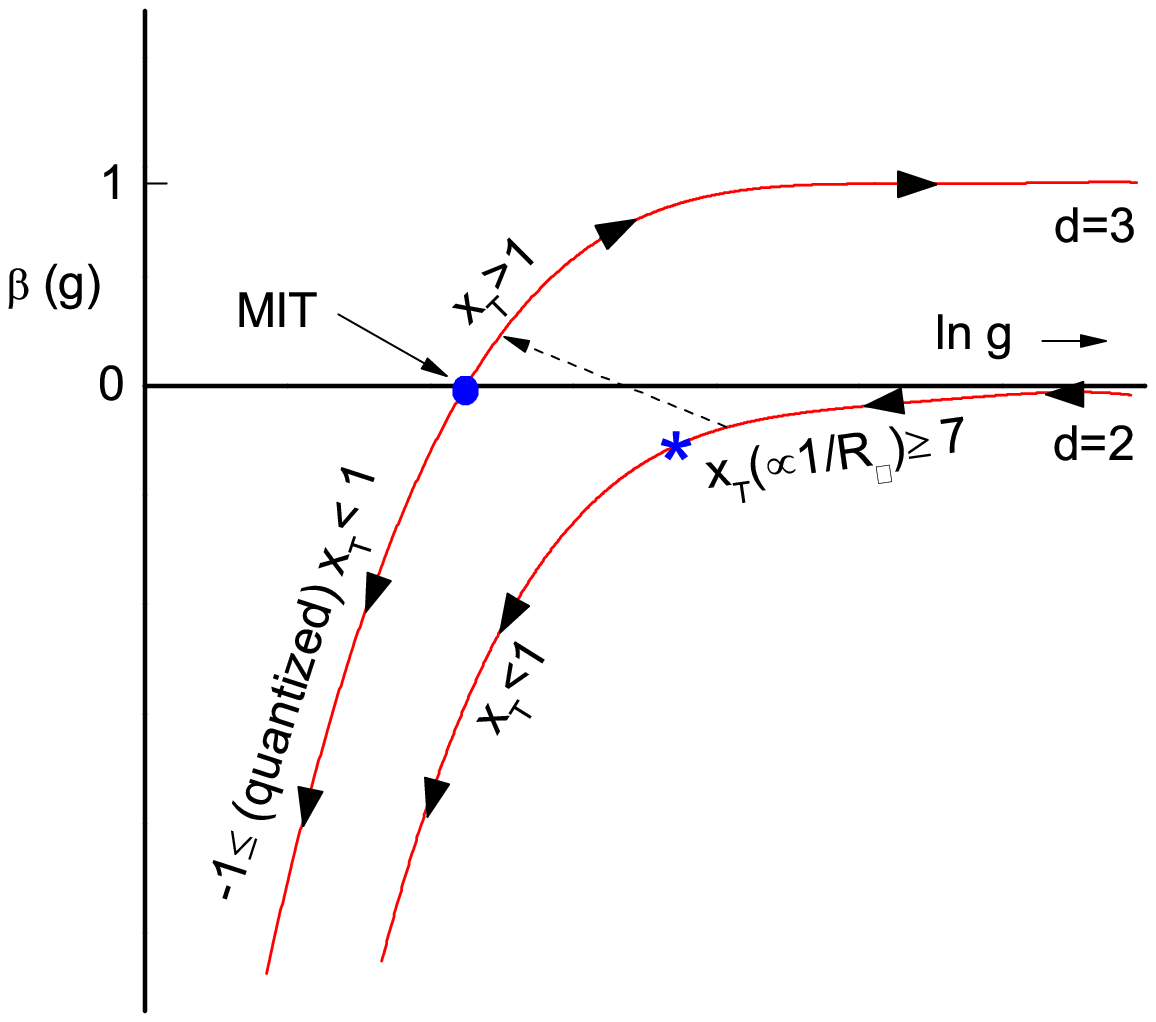}
\caption{Summary of the nonlinearity exponent $x_T$ in various regimes in the \textit{T}=0 scaling diagram of Abrahams et al.\cite {abrahams79}. The solid point represents the metal-insulator transition (MIT). The star on the 2D curve (corresponding to $R_ {\square} \approx 30 {\rm k}\Omega$) represents the WL-SL crossover. Both the MIT and WL-SL crossover are accompanied by the change in the exponent value: $x_T > 1$ on the metallic side and $x_T < 1$ on the insulation side. The dashed line schematically denotes crossover from 2D to 3D with increase of thickness.}
\label{fig.11}
\end{figure}

\begin{table*}
\caption{ The nonlinearity exponent $x_M$ for different materials (mostly 3D and exhibiting VRH) obtained using the scaling formalism. T, D and B represent temperature, disorder and magnetic field respectively as the control parameter. \textit{m} and $T_o$ are VRH parameters (Eq. 1) and the type refers to the scaling behavior (see text for details). $n$ is the quantization integer. Entries within various classes of materials are arranged chronologically. The error in an exponent is estimated to be about 10\% including errors in digitizing the source data except as noted. }

\begin{ruledtabular}
\begin{tabular}{llcclllrr} 

Systems          &$~m$ &$T_o$(K) &Type  &$~~~x_T$ &$~~x_D$ &$~~x_B$ &~~~~\textit{n}     &Data Source\\\hline  
\textit{a}-Semicond. \\
\textit{a}-SiO   &1/4 &$3.2\times 10^9$ &I &~0.09 	& & &1 		&A Servini et al., Thin Sol Films 3, 341 (1969) \\ \cline{1-8}
\textit{a}-Ge    &1/4  &$5\times 10^6$ &I &~0.24  	& & &3 		&N Croitoru et al., Thin Sol Films 3, 269 (1969) \\
\textit{a}-Ge    &1/4 &$7.9\times 10^7$ &I &~0.16 	& & &2 		&Ref. [\onlinecite {morgan71}] (1971)   \\
\textit{a}-Ge    &1/4  &$8.9\times 10^7$ &I &~0.16  	& & &2 		&M Telnic et al., Phys Stat Solidi(b) 59, 699 (1973) \\
\textit{a}-Ge    &1/4  &$1.4\times 10^8$ &I &~0.24 	& & &3 		&P J Elliot et  al., AIP Conf. Proc. 20, 311 (1974) \\
\textit{a}-Ge    &1/4  &$2.6\times 10^7$ &I &~0.23  	& & &3 		&T Suzuki et al., J Non-Cryst Solids 23, 1 (1977) \\ \cline{1-8}
\textit{a}-Ge:Cu &1/2	&86 &II &~0.38		& & &5	  	&A N Aleshin et al., Sov Phys Semicond. 21, 466(1987) \\
\textit{a}-Si:H:P&1  &1487  &I &~0.08 		& & &1 		&Ref. [\onlinecite {nebel92}] (1992) \\
\textit{a}-Si:H:P&1  &1487  &I & 		&~~0 & &0 		&Ref. [\onlinecite {nebel92}] (1992) \\
\textit{a}-Si:Y  & 1/2 &257 &I &~0.16\footnote[1]{$0.159 \pm 0.006$ obtained using the original data.}   & & &2 		&Ref. [\onlinecite {ladieu00}] (2000) \\
\textit{a}-CN:H  &1/4 &$2.9\times 10^8$ &I &~0.16  	& & &2 		&S Kumar et al., J Non-Cryst Solids  338-340,349 (2004) \\ \hline 
d-semicond. \\
\textit{n}-GaAs  &1/2 &110 &II &~0.31 	  & & &4 		&D Redfield, Adv Phys 24, 463 (1975) \\
\textit{n}-Si:P  &1/4 &$10^6$ &I &~0.13  	& & &2 		&Ref. [\onlinecite {rosen80}] (1980) \\
\textit{n}-Si:Mn & 1/2 &1000 &II &~0.15  	& & &2 		&A  V  Dvurechenskii et al., JETP Lett. 48, 155 (1988) \\
\textit{n}-GaAs  & 1/2 &9.4 &II &~0.17 	  & & &2          	&F Tremblay et al.,  Phys Rev B 40, 3387 (1989) \\
\textit{d}-Ge:Ga &1/2 &132 &I &~0.48		  & & &6		&T W Kenny et al., Phys Rev B 39, 8476 (1989) \\
\textit{n}-Zn:Se  &1/2 &400 &I &~0.31  	& & &4 		&I N Timchenko et al., Sov Phys Semicond. 23, 240 (1989) \\ \cline{1-8} 
\textit{n}-Si:As & 1/4 &$2.4 \times 10^7$ &I &~0.08	 	& & &1 		&C Gang et al., Solid State Comm 72, 173 (1989) \\
\textit{n}-Si:As &1/4 &760 &I &~0.30 	  & & &4 	        &R W van der Heijden et al., Phil Mag B 65, 849 (1992) \\ \cline{1-8}
\textit{p}-Si:B  &1/4 &$2\times 10^6$ &I &~0.24 	  & & &3 		&Y Shwarts et al., Sem Phy, Quan Elec  Opto 3, 400 (2000) \\
\textit{n}-CdSe  &1/2 &5200 &II &~0.16 	  & & &2 		&D Yu et al., Phys Rev Lett 92, 216802 (2004) \\
\textit{n}-Si:P:B  &1/2 &8.6 &I &~0  	& & &0 		&Ref. [\onlinecite {galea07}] (2007) \\ \hline 
Composite \\
Ni$_{0.24}$SiO$_{2~0.76}$  		 &1/2 &$2.3\times 10^4$ &I &~0.14 &          &  &2 	        &Ref. [\onlinecite {abeles75}] (1975) \\
C-PVC  		 &2/3 &112 &I &~0.63 &          &  &8 	        &L J Adriaanse et al., Synth Metals 84, 871 (1997) \\
SCNT-PMMA        &1/2 &1000 &I &~0.39 	 	& & &5 		&J M Benoit et al., Phys Rev B 65, 241405 (2002)  \\ \hline 
Dbl. perovskite \\
Ba$_2$MnReO$_6$  &1/2 &$3.8\times 10^4$ &I &~0.16 	& & &2 	        &B Fisher et al., J Appl Phys 104, 033716 (2008)\\ \hline
Metal cluster \\
Pd$_{561}$Phen$_{37}$O$_{200}$  &1/2 &$4096$ &I &~0.27 	& & &3 	        &Ref. [\onlinecite {staveren91}] (1991)\\ \hline
Organic crystal \\
$\theta$-(BEDT-TTF)$_2$ &~1 &24 &I &~0.14 & & &2 	        &Ref. [\onlinecite {takahide06}] (2006) \\
CsZn(SCN)$_4$ \\
$\theta$-(BEDT-TTF)$_2$ &~1 &24 &I &  	 & &~~0 &0 	        &Ref. [\onlinecite {takahide06}] (2006) \\
CsZn(SCN)$_4$ \\ \hline
2DEG			  &0.7 &1.6 &I  &~0.08 	& & &1 	        &Ref. [\onlinecite {gershen00}] (2008)\\

\end{tabular} 
\end{ruledtabular}
\end{table*}

Some points are in order. i) All values of the exponent determined so far lie within the bounds -1 and +1 respecting the lower bound argued earlier (Section II);  ii) The inset in Fig. 8a shows a histogram of the exponents in the panel (\textit{a}). The distribution exhibits a pronounced peak at \textit{n}=2 and is clearly asymmetric around the peak with a tail towards large \textit{n}; iii) For $x_M=0$, the field-dependence and dependence on the variable \textit{M} of conductance becomes completely separate (Eq. \ref{eq:scaling}). In particular, when \textit{M} is temperature  the field-dependence and the temperature-dependence become separated. This ia in contradiction to field-effect models (Eq. \ref{eq:lowF}); iv) The field scales in the WL regimes have been predicted earlier to remain unaffected by an applied magnetic field i.e. $x_B=0$ (Section III.A). We are not aware of any experiment to verify this. On the other hand, there are some data (Tables I and II) to suggest that $x_B$ may be also zero in 3D. In fact, the physical argument in support of the zero exponent given earlier is independent of dimensionality. Although a magnetic field does not impart any energy to electrons the zero value of $x_B$ is still significant because a magnetic field destroys time reversal invariance and is known to influence wave functions.

Fig. 10 summarizes the experimental findings about the nonlinearity exponent $x_T$ in disordered systems in 2D and 3D on the \textit{T}=0 scaling diagram\cite {abrahams79}. For highly conducting 3D samples, the nonlinear exponent is undefined. As the MIT is approached from the metallic side, $x_T > 1$. As MIT is just crossed to the insulating side, $x_T < 1$. Therefore, the MIT is also the point where $x_T$ appears to undergo a sharp transition in value from one $> 1$ to one $< 1$. For large conductance in 2D, the exponent $x_T$ is also large. As disorder increases it decreases and reaches an experimental (theoretical) limit of about 7 (3) at the WL-SL crossover marked by a star in the figure. In the SL regime in 2D a single experimental point indicates that $x_T <1$ as in 3D. The dashed line represents schematically a possible path corresponding to increasing thickness of a 2D film when the exponent decreases from a large value to one closer to 1 as discussed earlier (Fig. 2). \\

\section{Summary}
Earlier the field-dependent conductance was measured in two very different physical systems - one VRH system of doped conducting polymers\cite {talukdar11} and another strongly correlated manganites\cite {talukdar12} and was found to obey a phenomenological one parameter scaling relation. Both belonged to SL regimes. In this work a lot of similar data from literature belonging to the full spectrum of disorder in 2D and 3D have been analyzed and found to obey the same universal scaling except in few SL cases where the samples seem to be better described by hot electron effects. Heuristic arguments are presented for dependence of the field scale only on the Ohmic conductivity, which is a significant departure from the current theories. An exact scaling function for the WL regimes is derived and has been shown to match perfectly the experimental data. The scaling analysis led to an interesting finding of a hitherto unknown scaling-nonscaling transition as a function of temperature in SL regimes in both 2D and 3D. An approximate scaling function has been suggested for the scaling phase of the transition. The associated nonlinearity exponent possesses a spectrum of values characteristic of various disordered regimes - from values greater than 1 in highly conducting samples to values less than 1 in highly insulating samples. Most significantly, the exponent is found to be quantized in strong localization pointing toward a new physics and necessity for theoretical efforts - $x_T \approx 0.08 n$ where \textit{n} is an integer and can be both positive and negative including zero, and tentatively, $x_D \approx 0.08 n$ where \textit{n} can be zero or only negative. In VRH systems there is a broad trend of the exponent $x_T$ decreasing with the characteristic temperature $T_o$. Results are compared with current theories and limitations are discussed.

\section{Acknowledgments}
Authors thankfully acknowledge discussions with Asok Sen and Bikas Chakrabarti. They acknowledge help provided by Biswajit Das in processing some data and are grateful to F. Ladieu and late D. L'Hote for sharing their experimental data with us.

\bibliography{nonohmic}

\begin{thebibliography}{46}
\bibitem [{pap(){\natexlab{b}}}]{paper26_4}%
  \BibitemOpen
  \href@noop {} {} ({\natexlab{b}}),\ \bibinfo {note} {this bound was
  mistakenly mentioned as -1/2 in Ref. [\onlinecite {talukdar11}].}\BibitemShut
  {Stop}%
\bibitem [{pap(){\natexlab{c}}}]{paper26_1}%
  \BibitemOpen
  \href@noop {} {} ({\natexlab{c}}),\ \bibinfo {note} {this physical
  requirement needs not be valid when the control parameter is disorder. This
  allows for negative values for the exponent $x_D$ as seen later.}\BibitemShut
  {Stop}%
\bibitem [{pap(){\natexlab{e}}}]{paper26_3}%
  \BibitemOpen
  \href@noop {} {} ({\natexlab{e}}),\ \bibinfo {note} {such a large value
  usually indicates an exponential variation. However, we stick with the
  power-law to maintain continuity with rest of the discussion.}\BibitemShut
  {Stop}%
\end{thebibliography}

\appendix
\section{Field-dependent conductance in weak localization}
The Ohmic conductance in the 2D WL regime is characterized by the logarithmic dependence on temperature
\begin{equation}
\Sigma_o(T) = \Sigma'_o(T') + { \Sigma_u \lambda_T } \ln ( T/T' ),
\label{eq:2dRT}
\end{equation}
where $\Sigma_u = e^2/ 2{\pi}^2 \hbar$, $\lambda_T = \alpha p$, and $\alpha$ an universal constant\cite {anderson79}. The exponent \textit{p} is defined by ${\tau}_{in}=a T^{-p}$ where ${\tau}_{in}$ is the inelastic relaxation time and \textit{a} is a constant. The logarithmic term in the above represents a small quantum correction to the classical conductance. The effects of applying an electric field are described by the hot electron model\cite {anderson79} where the system is assumed to absorb all the energy imparted by the field giving rise to an effective carrier temperature $T_e$. The latter is given by the energy balance relation:
\begin{equation}
\Sigma(V) V^2 {\tau}_{in}(T_e) = {1 \over 2} \gamma W ({T_e}^2 - T^2),
\label{eq:hem}
\end{equation}
or,
\begin{equation}
(2a/\gamma W )\Sigma(V) V^2 + T^2 {T_e}^p = {T_e}^{2+p}, \nonumber
\end{equation}
where $\gamma T$ is the electronic specific heat and $W$ is the volume of the sample. As a first approximation one can put $T_e \approx T$ and $\Sigma \approx \Sigma_{\square o}$, the temperature-independent sheet conductance on the left hand side of the above equation and obtain
\begin{equation}
{ (T_e}/T )^{2+p}= 1 + {(V/V_o)}^2,
\label{eq:Te}
\end{equation}
where $V_o$ is given by
\begin{equation}
V_o^2 = { \gamma W T^{2+p} \over 2a \Sigma_{\square o}}.
\label{eq:Afscale}
\end{equation}
Using Eq. \ref{eq:Te} in Eq. \ref{eq:2dRT} gives
\begin{equation}
\Phi_{WL} = 1 + {z_F \over 2} \ln (1 + q^2 ),
\label{eq:lnRV}
\end{equation}
where $ \Phi_{WL} = {\Sigma(T,V) / \Sigma_o(T)} $, $q = V / V_o$, $z_F = \Sigma_u \lambda_V /  \Sigma_{\square o} $ and $ \lambda_V = \lambda_T /(1+p/2)$.

\section{Field-dependent resistance in the scaling phase of the SNST}
The HEM-SL does not generally support scaling (Section II.B) and hence, is not supposed to describe the \textit{R-V} data particularly in the scaling phase. However the relevant data at $T \leq T_c$ in Ref. \onlinecite {rosen80} seem to be well described by the model in contrast to the disagreement in Ref. \onlinecite {gershen00}. Here we show that the HEM-SL does support scaling under the special condition that the voltage scale $V_o$ is a constant corresponding to $x_T = 0$ as is the case in Ref. \onlinecite {rosen80} (Fig. 6).  Eq (\ref{eq:hemfs}) gives the voltage scale in the HEM-SL: $V_o^2 = C R_o T^{\beta} $ (up to a scale factor) so that
\begin{equation}
R_o = V_o^2 /C T^{\beta},
\label{eq:hemRo}
\end{equation}
which is numerically equivalent to Eq. (\ref{eq:mott}). The field-dependent resistance is then obtained by replacing \textit{T} with the effective temperature $T_e$. Putting the latter from Eq. (\ref{eq:hemSL}) into the above equation yields the simple scaling expression for \textit{R}:
\begin{equation}
R/R_o = 1 - {(V / V_o)}^2.
\end{equation}
The scaling function $\Phi_H = R_o/R$ can be written as
\begin{equation}
{\Phi}_H^{-1} = 1 - q^2,
\label{eq:hemsf0}
\end{equation}
for $q =V / V_o \leq 1$.

We now consider a situation when $V_o$ in the scaling phase is not a constant and is given by an additional expression $V_o = a_T R_o^{-x_T}$ where $a_T$ is a constant. An example is the 2DEG discussed in Section III.B where $x_T \approx 0.08$ (Fig. 4). For considerations within the HEM-SL we ignore the deviation from Eq. (\ref{eq:hemRo}) (compare insets of Figs. 4 and 5) and obtain
\begin{equation}
R_o^{1+2x_T} = a_T^2 /C T^ {\beta}.
\end{equation}
The above is valid only for $x_T > -1/2$. Greater $x_T$ is, larger will be the deviation. Proceeding as before finally gives the following transcendental equation for $\Phi $:
\begin{equation}
{\Phi}_H^{2x_T} - {\Phi}_H^{-1} = q^2.
\label{eq:hemsf}
\end{equation}
The scaling function above is remarkable in that it contains no other parameter than $x_T$ which of course remains undetermined. For $x_T>0$, it clearly satisfies the SCR (Eq. \ref{eq:zT}). For large \textit{q}, $\Phi_H \sim q^{1/x_T}$ so that $z_F = 1/x_T$. For $x_T=0$, Eq. (\ref{eq:hemsf}) reduces to Eq. (\ref{eq:hemsf0}).

\section{Nonlinearity exponents in other systems}
\begin{table}[h]
\caption{ The nonlinearity exponents for different conducting polymers. Symbols are same as in Table I.}
\begin{ruledtabular}
\begin{tabular}{lccccccr}

System   &\textit{m} &$T_o$(K) &Type  &$x_T$     &~$x_D$  &~$x_B$    & ~\textit{n}     \\
\hline

PPy(powd.)\footnote[1]{Ref. [\onlinecite{talukdar11}]}  	&1/4 &$1.7\times 10^6$ &I  &-0.33$\pm$0.01  & 			 & 	   &~-4  \\
PPy(film)\footnotemark[1]   	  &1/3 &2700 &I  &-0.16$\pm$0.01  & 			 & 	   &~-2  \\
PPy(film)\footnotemark[1]   	  &1/2 &2340 &I  &~0.23$\pm$0.01  & 			 & 	   &~~3  \\
PPy(film)\footnote[2]{Ref. [\onlinecite{varade13}]}   	  &1/4 &10674  &I  &~0.55$\pm$0.03                   &        &     &~~7  \\
PPy(film)\footnotemark[2]  &1/4 &82  &I  &~0.74$\pm$0.03  &        &     &~~9  \\ \cline{1-5}
PPy(film)\footnotemark[1]   	  &1/2 &2340 &I  &                   &~-0.3 &     &~-4  \\
PPy(film)\footnotemark[1]   	  &1/2 &2340 &I  &                   &        &~0 &~~0  \\
PEDOT\footnotemark[1]  				  &1/2 &1798 &I  &~0.16$\pm$0.01  & 			 & 	   &~~2  \\
PDA\footnote[3]{Quasi-1D crystals (Aleshin et al., Phys. Rev. B, {\bf 69}, 214203 (2004)) }				      &2/3 &1265 &II &~0.50$\pm$0.03\footnotemark[1]  &        &     &~~6  \\

\end{tabular} 
\end{ruledtabular}
\end{table}

\begin{table}
\caption{ The nonlinearity exponents for CMR phases in different manganites (Ref. \onlinecite{talukdar12}). PI and FM represents the paramagnetic insulating and ferromagnetic phases.}
\begin{ruledtabular}
\begin{tabular}{lcccc}

System      &~~~~\textrm{$x^{FM}_T$}  &~~\textit{$n^{FM}$}   &~~~\textrm{$x^{PI}_T$}  &~~\textit{$n^{PI}$}      \\ \hline

${\rm Sm}_{0.55}({\rm Sr}_{0.5}{\rm Ca} _{0.5})_{0.45}\rm{MnO}_3$\footnote{Single crystal}  & &  & ~0 &~0 \\
${\rm Sm}_{0.55}{\rm Sr}_{0.3375}{\rm Ca}_{0.1125}{\rm MnO}_3$   & $~~0.17 $ & ~~2  &$~~0.09$ &~1\\
${\rm Sm}_{0.55}{\rm Sr}_{0.45}{\rm MnO}_3$	& $-0.23$ &~-3 &$-0.14$ &-2 \\
${\rm La}_{0.275}{\rm Pr}_{0.35}{\rm Ca}_{0.375}{\rm MnO}_3$     & $-0.09 $ &~-1 &~~0 &~0\\
${\rm La}_{0.87}({\rm Mn}_2{\rm O}_3)_{0.13}{\rm MnO}_3$ &  $-0.83$ &-10 & $-0.16$ &-2\\
${\rm La}_{0.75}{\rm Ca}_{0.25}{\rm MnO}_3$    &  $~~0.27$  &~~3 & $~~0.27 $ &~3\\
${\rm La}_{0.75}{\rm Ca}_{0.25}{\rm MnO}_3/{\rm BaTiO}_3$  & $-0.70$ &~-9  & $-0.15$ &-2 \\ 
\end{tabular} 
\end{ruledtabular}
\end{table}

\end{document}